\documentclass[aps,12pt,showpacs,nofootinbib,amsmath, floatfix]{revtex4}
\usepackage{amsmath}
\usepackage{amsfonts}
\usepackage{amssymb}
\usepackage{color}

\def\bc{\begin{center}}
\def\nno{\nonumber}
\def\ec{\end{center}}
\def\be{\begin{eqnarray}}
\def\ee{\end{eqnarray}}

\newcommand{\omits}[1]{}

\definecolor{dyellow}{rgb}{1.,0.8,.0}
\definecolor{myblue}{rgb}{.1,.1,.7}
\definecolor{dcyan}{rgb}{.0,.6,.6}
\definecolor{dmagenta}{rgb}{0.6,0.0,0.6}
\definecolor{brown}{rgb}{0.6,0.2,0.}
\definecolor{darkblue}{rgb}{.0,.0,0.5}
\definecolor{darkred}{rgb}{0.75,0.0,0.0}
\definecolor{orange}{rgb}{1.,.6,.0}
\definecolor{dorange}{rgb}{0.8,.4,.0}
\definecolor{darkgreen}{rgb}{0.0,0.6,0.0}
\definecolor{purple}{rgb}{.4,.0,.4}

\def\ka{\kappa}
\def\la{\lambda}

\def\d#1#2{\frac{\displaystyle #1}{\displaystyle #2}}
\def\r{\partial}

\newcommand{\UWF}{Umov-Weyl-Fock~}
\newcommand{\UWFL}{Umov-Weyl-Fock-Lorentz~}
\newcommand{\dS}{$d{S}$}
\newcommand{\AdS}{${A}d{S}$}
\newcommand{\Mink}{$Mink$}

\newcommand{\PoR}{principle of relativity}

\newcommand{\SR}{special relativity}

\newcommand{\LFT}{${LFT}$}

\newcommand\btd{\raise 2pt
\hbox{$\hat\bigtriangledown$}\hskip 1.5pt}
\newcommand\bt{\raise 2pt
\hbox{$\bigtriangledown$}\hskip 1.5pt}

\def\PRD{{\it Phys. Rev.}~{\bf D}}

\def\PLA{{\it Phys. Lett.}~{\bf A}}
\def\PLB{{\it Phys. Lett.}~{\bf B}}
\def\GRG{{\it Gen. Rel. Grav. }}

\def\CTP{{\it Commun. Theor. Phys. }}

\begin{document}

\title{The Principle of Relativity and the Special Relativity Triple}%

\author{{Han-Ying Guo}$^{1}$}
\email{hyguo@itp.ac.cn}
\author{{Hong-Tu Wu}$^{2}$}
\email{lobby_wu@yahoo.com.cn}
\author{{Bin Zhou}$^{3}$} \email{zhoub@bnu.edu.cn}

\affiliation{%
${}^1$ Institute of Theoretical Physics, Chinese Academy of
Sciences, Beijing 100190, China,}
\affiliation{%
${}^2$Department of Mathematics, Capital Normal University, Beijing
100048, China,}
\affiliation{%
${}^3$Department of Physics, Beijing Normal University, Beijing
100875, China.}

\date{December 2008}

\begin{abstract}
Based on  the principle of relativity and the postulate on
universal invariant constants ($c,l$) as well as Einstein's
isotropy conditions,  three kinds of special relativity  form a
triple with a common Lorentz group as isotropy group under full
Umov-Weyl-Fock-Lorentz transformations among inertial motions.
\end{abstract}


\maketitle

\tableofcontents

\section{Introduction}

 Recent observations in
precise cosmology show that our universe is accelerated expanding
and quite possibly asymptotic to a Robertson-Walker-like de Sitter
(\dS) spacetime with a tiny cosmological constant $\Lambda$,
rather than the flat Minkowski (\Mink) spacetime \cite{SN,WMAP}.
With lots of puzzles, these greatly challenge Einstein's theory of
relativity as a foundation of the cosmic scale physics
characterized by $\Lambda$. In order to face the challenges, it
would be better to re-examine the principles of Einstein's theory
from the very beginning.

As is well known, like the Galilei \PoR\ in Newtonian mechanics
the Poincar\'e \PoR\  plays the fundamental role  in  Einstein's
\SR. It requires: {\it For a set of inertial reference frames, in
which free particles including light signals move with uniform
velocities along straight lines, the laws of nature hold good in
their simplest
 form invariant under the transformations among the frames} (see,
e.g. \cite{Poincare, 1905}). There is a simple and important problem
for the principle: What are the most general transformations among
the inertial motions in the inertial frames?

Long ago, Umov \cite{Umov}, Weyl \cite{Weyl} and Fock \cite{Fock}
investigated this problem. It turns out that those transformations
are  linear fractional ones with common denominators (\LFT s),
called the Umov-Weyl-Fock-transformations denoted by $\cal T$.
Since all physical coordinates should have right dimensions in
general, in order to make the time coordinate has the same
dimension with three spacial ones, there should be two universal
invariant constants: the speed of signal $c$ and the length $l$,
which are universal for all inertial frames and invariant under
their transformations. Thus, there should be a postulate on
universal invariant constants $(c,l)$ for the transformations
among the inertial frames, although in Galilei transformations
they  do not appear explicitly and neither does $l$ in Poincar\'e
transformations. Clearly, these transformations form a Lie group
with twenty four parameters as the transformation group among
inertial motions in inertial frames denoted by ${IM}(4)$, whose
algebra $\mathfrak{im}(4)$ may be called the inertial motion
algebra. In previous investigations it is immediately turned  to
the transformations of Poincar\'e group $ISO(1,3)$  on the
\Mink-spacetime. Such a group $IM(4)$ that contains the Lorentz
group $SO(1,3)$ as isotropy group is denoted by $IM_L(4)$ and its
transformations called the
 {\it  Umov-Weyl-Fock-Lorentz transformations} ${\cal T}_L$.
However, there are ten generators for the Poincar\'e
transformations
only, which  form merely a subset of %
${\cal T}_L$. What is the role played by the rest transformations
of $IM_L(4)$? This problem
  had been ignored for long time.

Later, the \dS/anti-de Sitter (\AdS)  \SR\ has been proposed
\cite{ Lu,LZG, Hua, psr, BdS, TdS,IWR,NH,Lu05, C3,Yan, OoI, PoI,
Lu80}, based on the \PoR\ \cite{Lu,LZG} and the postulate on two
universal constants ($c$ and $l$)\cite{BdS}. The \dS/\AdS\, \SR\
is invariant under the \LFT s of \dS/\AdS-group $SO(1,4)/SO(2,3)$
among the inertial motions in the Beltrami model \cite{beltrami}
of \dS/\AdS-spacetime \cite{BdS,Lu80} with curvature radius $l$,
respectively. Since there is also  a Lorentz group
 as  isotropy group in both,  the
\LFT s of  \dS/\AdS-group are also  subsets of the
 ${\cal T}_L$. However, there
are  thirty parameters for three kinds of \SR\ in total, this seems
another puzzle. How to describe them at the same time in view of the
$IM_L(4)$?

In this Letter, we show that three kinds of special relativity
form such a \SR\ triple in $IM_L(4)$ that there are a common
isotropy Lorentz group and some of the \UWFL\ transformations
among them under $IM_L(4)$.

These can be shown  by means of the algebraic analysis and
specially the linear combinations in the Lie algebraic space of
$\mathfrak{im}_L(4)$. In fact, if the
  Lorentz algebra $\mathfrak{so}(1,3)$ as the isotropy algebra,
 the
Poincar\'e/\dS/\AdS-algebra
$\mathfrak{iso}(1,3)/\mathfrak{so}(1,4)/\mathfrak{so}(2,3)$ as
subalgebras simultaneously  should share the common isotropy
algebra $\mathfrak{so}(1,3)$. Further, the translation
 generators of  Poincar\'e  algebra
$\mathfrak{iso}(1,3)$ can be given by the linear combination
(plus) of the `translations' generators of \dS/\AdS-algebra
$\mathfrak{so}(1,4)/\mathfrak{so}(2,3)$ in the space of
$\mathfrak{im}_L(4)$. And rest generators including four of Cartan
subalgebra $\mathfrak h$ transform either the `translations' in
\dS/\AdS-algebra $\mathfrak{so}(1,4)/\mathfrak{so}(2,3)$ from each
other or the translations in  Poincar\'e algebras
$\mathfrak{iso}(1,3)$ among themselves. In this sense, three
subalgebras $\mathfrak{iso}(1,3),\mathfrak{so}(1,4)$, and
$\mathfrak{so}(2,3)$ as well as their groups form a triple under
$IM_L(4)$. Since these symmetries correspond to  three kinds of
\SR, respectively, three kinds of \SR\ do  form the \SR\ triple
under
 ${\cal T}_L$.  In
addition, the  linear combination (minus) of \dS/\AdS's
`translation' operators may  lead to another set of four
generators that with generators of $\mathfrak{so}(1,3)$ forms the
second Poincar\'e algebra and whereas the linear combinations of
these generators of two Poincar\'e algebras may also lead to the
\dS/\AdS\ algebra, too. With the help of the linear combinatory in
the space of $\mathfrak{im}_L(4)$, which is different from
In\"on\"u-Wigner's contraction procedure\cite{IW},
 we would like to show the \SR\
triple step by step more physically in this Letter.

This Letter is arranged as follows. In section \ref{sec: UWF}, we
recall the \UWF transformations ${\cal T}$ for the \PoR\ {with two
universal invariant constants $(c,l)$} and how to
 get the symmetries of three kinds of \SR. In  section \ref{sec: SRT}, we
show the linear combinatory relations between the `translations'
of \dS/\AdS\ algebra and those of two Poincar\'e algebras. And we
also show that under the \UWFL transformations ${\cal T}_L$  three
kinds of \SR\ do form the \SR\ triple. Finally, we end with some
remarks.


\section{The Principle of Relativity and
Three Kinds of Special Relativity}\label{sec: UWF}

\subsection{The Principle of Relativity and  Most General Transformations for Inertial Motions}

For the \PoR, in the inertial coordinate  frames  ${\cal
F}:=\{S(x)\}$, a free particle takes inertial motion described by%
 \be\label{eq:uvm}%
 x^i=x_0^i+v^i(t-t_0),~~
v^i=\frac{dx^i}{dt}={\rm consts}.~~ i=1, 2, 3. %
\ee%
What are the most general transformations ${\cal T}:=\{T\}$
\be\label{eq:FL}%
{\cal T}\ni T:\quad {x'}^\mu=f^\mu( x, T), ~ x^0=ct, ~ \mu=0,\cdots, 3,%
\ee%
that keep  Eq. (\ref{eq:uvm})
invariant? Here, $c$ is  invariant under ${\cal T}$.%

It can be  proved \cite{Umov, Weyl,Fock} that {\it the most general
form of the transformations (\ref{eq:FL}),
 which transform a
uniform-velocity, straight-line motion (\ref{eq:uvm}) in $S(x)$ to a
motion of the same nature in $S'(x')$, are that the four functions
$f^\mu$ are ratios of linear functions, all with the same
denominators, i.e. the \LFT s}
\begin{equation}
{ T}:\quad  l^{-1}x'^\mu = \frac{A^\mu_{\ \nu} l^{-1} x^\nu +
b^\mu}{c_\lambda  l^{-1}x^\lambda + d} \label{eq:LFT}
\end{equation}
with
\begin{equation}\nno
  det\ { T}=\left| \begin{array}{rrcrr}
    A    & b^t \\
    c
     & d
  \end{array} \right|
  =1, 
\label{eq:det}
\end{equation}
 where $l$ is another
 universal length  parameter invariant under  $T\in \cal
 T$ and $A=(A^\mu _{~\nu})$ is a $4\times 4$ matrix, $b,c$  $1\times 4$ matrixes, $d\in R$,
  up-index
  $\small{^ t}$
 denotes the transpose.
Thus, together with the \PoR, there is a postulate on universal
invariant constants:  {\it For all the inertial frames, there are
two universal constants $(c,l)$ with dimension of speed and length
invariant under ${\cal T}$, respectively}.
  A simpler proof for  Umov-Weyl-Fock
transformations is given in  appendix of \cite{duality}.

It is clear that  an Umov-Weyl-Fock transformation $T\in {\cal T}$
(\ref{eq:LFT}) is represented by a non-singular $5\times5$
 matrix with
twenty four independent entries, all these matrices $\forall T\in
\cal T$ form the  inertial motion transformation group
$IM(4)$ with generators  defined by%
\be\label{eq: gntrs}%
T_a:=\frac{\partial {x'}^\mu(x, \tau^a) }{\partial
\tau^a}|_{\tau^a=e}\frac{\partial}{\partial x^\mu},\quad \tau^a\in {T},\quad a=1,\cdots, 24,%
\ee%
forming the inertial motion algebra $\mathfrak{im}(4)$ as Lie
algebra of  $IM(4)$. Actually, the inertial motion (\ref{eq:uvm})
may be viewed as a straight line and the most general
transformations among straight lines may form a group, which may
be referred to the real projective transformations of the real
projective group ${RP}(4)$(see, e.g. \cite{psr, Hua}). But, care
should be taken for the orientation \cite{BdS}.

 However,
 Umov, Weyl and Fock did not go further rather
immediately went back to Poincar\'e transformations by requiring,
say,  the plane wave front equation of light to be invariant
\cite{Fock}.

\subsection{From Einstein's Isotropy Conditions to  Einstein's Special Relativity}

As is well known, the \PoR\ connotatively assume that there should
be
 isotropy and homogeneity for all points in a given frame $S(x)$. We may first consider the
 isotropy by studying the inertial motions passing through the origin
$O({o}^\mu=0)$ of  $S(x)$.
 This  leads to the Lorentz
group as isotropy group at the origin. Then we consider the
homogeneous property among all points by studying how to transit
the origin $O(o^\mu=0)$ to the other  point $A(\forall a^\mu\neq
0)$ in $S(x)$ such that the point $A$ becomes the origin
$O'(o'^\mu=0)$ in the transformed frame $S'(x')$ by some
transformations generated by other four generators of  ${\cal
T}_L$.

Let us  consider the isotropy first as Einstein did in 1905.

{\it ``At the time $t=\tau=0$, when the origin of the co-ordinates
is common to the two frames, let a spherical wave be emitted
therefrom, and be propagated with the velocity $c$ in system $K$. If
$(x, y, z)$ be a point just attained by this wave, then
$$x^2+y^2+z^2=c^2t^2.$$
 Transforming this equation with the aid of
our equations of transformation we obtain after a simple calculation
$$\xi^2+\eta^2+\zeta^2=c^2\tau^2$$
The wave under consideration is therefore no less a spherical wave
with velocity of propagation $c$ when viewed in the moving system.
This shows that our two fundamental principles are compatible.
\footnote{The equations of the Lorentz transformation may be more
simply deduced directly from the condition that in virtue of those
equations the relation $x^2+y^2+z^2=c^2t^2$ shall have as its
consequence the second relation
$\xi^2+\eta^2+\zeta^2=c^2\tau^2$.}"} \cite{1905} Actually,
Einstein considered the motions of light signals passing though
the common origin of the inertial frames, and introduced the
light-cone
at the origin may be called {\it Einstein's light-cone}%
 \be\label{eq:Elightcone}%
{\cal D}_0:&& \eta_{\mu\nu}\triangle x^\mu \triangle x^\nu|_O:=\eta_{\mu\nu}x^\mu x^\nu|_O=0, ~~ x^0=ct,~\mu,\nu=0,\cdots,3,\\\nno %
&&J:=(\eta_{\mu\nu})_{\mu,\nu}=diag (1,-1,-1,-1),
 \ee%
where, $\triangle x^\mu:=(0, x^\mu)$. For an infinitesimal time
duration  and  spacial intervals  $dx^\mu=lim_{\triangle x^\mu\to
0}\triangle x^\mu$, it follows the infinitesimal form of Einstein's
light-cone
\be\label{eq:ds0}%
{\cal D}_{0,dx}:\quad ds^2|_{O}:=\eta_{\mu\nu}dx^\mu dx^\nu|_O=0.%
\ee%

For other free massive particles or ``tachyons" passing through the origin with uniform
velocity%
\be%
v^i:=\frac{dx^i}{dt}=consts,
\ee%
they should satisfy $x^2+y^2+z^2=v^2t^2\lessgtr c^2t^2$, respectively. Thus,  %
the time-like/null/space-like condition at the origin
or  its infinitesimal form reads, respectively %
\be\label{eq:Econditions}%
 \eta_{\mu\nu}x^\mu x^\nu |_O\gtreqless0,\qquad ds^2|_O=\eta_{\mu\nu}dx^\mu dx^\nu|_O
 \gtreqless0,
 \ee%
which may be called {\it Einstein's isotropy conditions} and {\it
Einstein's metric}, respectively.

Actually,  from the invariance of  Einstein's isotropy conditions
and  Einstein's metric (\ref{eq:Econditions}) under the
transformations (\ref{eq:LFT}), it follows  a set of
transformations $L^{\mu'}_{\ \mu}\subset A^{\mu'}_{\ \mu} $ of
Lorentz group as isotropy group
\be\label{eq:SO(1,3)}%
{\cal L}:\quad \eta_{\mu\nu}=\eta_{\mu'\nu'}L^{\mu'}_{\
\mu}L^{\nu'}_{\ \nu}, \quad L^{\mu'}_{\ \mu}\in SO(1,3)\equiv {\cal L} \subset \cal T.%
\ee%
Then,   $\cal T$ should be the Umov-Weyl-Fock-Lorentz
transformations ${\cal T}_L$. Up to now, there are no restrictions
on the rest eighteen parameters of
 ${\cal T}_L$.

 Einstein also required: ``{\it
The laws by which the states of physical systems undergo change are
not affected, whether these changes of state be referred to the one
or the other of two systems of co-ordinates in uniform translatory
motion.}"\cite{1905}  This means that the  4-d translation group
$R(1,3)$ holds for the transitivity. From this requirement, it
follows Poincar\'e transformations of
 $ISO(1,3)$ as the semidirect product of  translations and %
 Lorentz transformations $R(1,3)\rtimes \cal L$%
\be\label{eq:ISO(1,3)}%
{\cal P}: \quad x^\mu \to  x'^\mu= L^\mu_{\ \nu} (x^\nu-a^\mu).%
\ee%
Both isotropy and homogeneity make the \Mink-spacetime as a 4-d
homogeneous space $M={\cal P}/{\cal L}$. And Einstein's isotropy
conditions (\ref{eq:Econditions}) hold
 simultaneously at all
points on it%
\be%
\eta_{\mu\nu}x^\mu x^\nu \gtreqless 0,\qquad ds^2=
\eta_{\mu\nu}dx^\mu
dx^\nu\gtreqless 0.%
\ee%

In view of the ${\cal T}_L$,  the Poincar\'e group is a subgroup of $IM_L(4)$ with  generators 
\be\label{eq:generator}%
  {\mathbf P}_\mu = \partial_\mu,~~
  x_\mu:=\eta_{\mu\nu}x^\nu,\qquad\\
  {\mathbf L}_{\mu\nu}
  = x_\mu \partial_\nu - x_\nu \partial_\mu \in \mathfrak{so}(1,3) \label{eq:Lorentz}
\ee%
 forming an $\mathfrak{iso}(1,3)$ algebra as a subalgebra of $\mathfrak{im}_L(4)$
\be\label{eq:iso13} \nno%
  [ \mathbf{P}_\mu, \mathbf{P}_\nu ] =0,\quad
  {[} \mathbf{L}_{\mu\nu},\mathbf{P}_\ka {]} =
    \eta_{\nu\ka} \mathbf{P}_\mu - \eta_{\mu\ka} \mathbf{P}_\nu,\quad\\
  {[} \mathbf{L}_{\mu\nu},\mathbf{L}_{\ka\la} {]} =
    \eta_{\nu\ka} \mathbf{L}_{\mu\la}
  - \eta_{\nu\la} \mathbf{L}_{\mu\ka}
  + \eta_{\mu\la} \mathbf{L}_{\nu\ka}
  - \eta_{\mu\ka} \mathbf{L}_{\nu\la}.
\ee%

\subsection{From Einstein's Isotropy Conditions to Other Two Kinds of Special Relativity}

In order to get  other subsets of inertial motions including Lorentz
transformations for  Einstein's isotropy conditions
(\ref{eq:Econditions}), it should be relaxed Einstein's assumptions
on Poincar\'e translations. This is also deeply motivated by recent
observations in precise cosmology \cite{SN,WMAP}.

In order to do so, we may first rewrite Eq. (\ref{eq:Elightcone}) to
get %
\be\label{eq:domains}%
{\cal D}_\pm:\quad \sigma_\pm(x):=\sigma_\pm(x, x)=1\mp l^{-2}\eta_{\mu\nu}x^\mu x^\nu >
 0,%
\ee%
 which denote two regions on  \Mink-spacetime with  boundary as `pseudosphere', respectively (see, e.g. \cite{Synge})%
\be\label{eq: sigma=0}%
\partial({\cal D}_\pm):\quad  \sigma_\pm(x)=1\mp l^{-2}\eta_{\mu\nu}x^\mu x^\nu = 0.%
 \ee%

 However, if the flatness of
 spacetime is relaxed, the conditions (\ref{eq:domains}) would
mean that the sets of all points in the frames $\{S(x)\}$ with
Einstein's isotropy conditions (\ref{eq:Econditions}) at the
origin, but without the Poincar\'e translations $R(1,3)$.
Therefore, we need to find the other  four generators for
transitivity. And together with the Lorentz isotropy group they
should also form a group with ten generators, respectively. It is
clear  both the \dS/\AdS-group should be the candidates.

Actually,  conditions (\ref{eq:domains}) without  flatness are
just the same as the domain conditions of inertial  frames in the
Beltrami model of \dS/\AdS-spacetime with radius $l$, respectively
\cite{BdS,Lu80}. If we regard the inertial coordinates $x^\mu$ as
the Beltrami coordinates in a chart $U_4$ \cite{BdS,Lu80}, say, of
the Beltrami atlas\footnote{In order to preserve the orientations,
the antipodal identifications should not be taken as was just
mentioned \cite{BdS,Lu80}.
}%
\be\label{eq:BLxU4}%
x^\mu=l\frac{\xi^\mu}{\xi^4},\quad \xi^4> 0,%
\ee%
the domain conditions (\ref{eq:domains}) become%
\be\label{eq:hyperboloids}%
 {\cal H}_\pm:\quad \eta_{\mu \nu}\xi^\mu \xi^\nu \mp (\xi^4)^2 \lessgtr 0.%
\ee%
These are just the \dS/\AdS-hyperboloid  ${\cal H}_\pm$,
respectively. %

From the Beltrami  model of  \dS/\AdS-spacetime \cite{BdS,Lu80},
it follows that there are the $LFT$s of  \dS/\AdS-group
$SO(1,4)/SO(2,3) \subset {\cal T}_L$,
\be\label{eq:SO14,23}\nno%
{\cal S}_\pm:\quad x^\mu\rightarrow {x'}^\mu&=&\pm
\sigma_\pm^{1/2}(a)\sigma_\pm^{-1}(a,x)(x^\nu-a^\nu)D_{\pm~\nu}^{~\mu},\\
D_{\pm~\nu}^{~\mu}&=&L_{~\nu}^\mu\pm l^{-2}%
\eta_{\nu \la}a^\la a^\ka
(\sigma_\pm(a)+\sigma_\pm^{1/2}(a))^{-1}L_{~\ka}^\mu,\\\nno
L&:=&(L_{~\nu}^\mu)\in SO(1,3),%
\ee
which transform a point $A(\forall a^{\mu}\neq0)$ with
$\sigma_\pm(a^{\mu})>0$ in the system $S(x)$  to the origin of  $
S'(x')$ and keep the domain conditions (\ref{eq:domains}) and the
Beltrami metrics invariant, respectively
\be\label{eq: BKmetric}%
ds_\pm^2=[\eta_{\mu\nu}\sigma_\pm^{-1}(x)\pm l^{-2}
\eta_{\mu\la}\eta_{\nu\ka}x^\la x^\ka
\sigma_\pm^{-2}(x)]dx^\mu dx^\nu. %
\ee%
 It is straightforward
to see that  at the origin of  frames, both  $ds^2_\pm|_O$ become
Einstein's metric in Eq. (\ref{eq:Econditions}). Therefore, the \LFT
s ${\cal S}_\pm$ of \dS/\AdS-group do form a subset of
Umov-Weyl-Fock-Lorentz transformations, ${\cal S}_\pm \subset {\cal
T}_L$, respectively. In fact, the domain conditions
(\ref{eq:domains}), the Beltrami-metrics (\ref{eq:
BKmetric}) and the light-cones \cite{BdS, Lu80}%
\be\label{eq:Blightcone}%
{\cal F}_{\pm}:=
\{\sigma_\pm(a,x) -[\sigma_\pm(a)\sigma_\pm(x)]^{1/2}\}\gtreqless0.%
 \ee%
 can be deduced by  \LFT s (\ref{eq:SO14,23}) from  Einstein's isotropy
 conditions  (\ref{eq:Econditions}), respectively. Further, due to the transitivity of
$LFT$s (\ref{eq:SO14,23}), the Beltrami-\dS/\AdS\ spacetime ${\cal
B}_\pm$ with the domain condition is also homogeneous, i.e. ${\cal
B}_\pm \cong {\cal S}_\pm/\cal L$, respectively. It is definitely
true globally chart by chart for the entire Beltrami-\dS/\AdS\
spacetime (see, e.g.\cite{BdS, Lu80}).

From the $LFT$s ${\cal S}_\pm$ (\ref{eq:SO14,23}), it is
straightforward to get their generators, which are also the Killing
vectors with respect to the Beltrami  metrics (\ref{eq: BKmetric})
up to some coefficients, as follows
\be\label{eq:generatorpm}%
  {\mathbf P}^\pm_\mu =(\delta_\mu^\nu\mp l^{-2}x_\mu x^\nu) \partial_\nu,~~
  x_\mu:=\eta_{\mu\nu}x^\nu,\qquad\\
  {\mathbf L}^\pm_{\mu\nu} = x_\mu {\mathbf P}^\pm_\nu - x_\nu {\mathbf P}^\pm_\mu
  = x_\mu \partial_\nu - x_\nu \partial_\mu ={\mathbf L}_{\mu\nu}\in \mathfrak{so}(1,3),
  \label{eq:Lorentz'}
\ee%
which form the $\mathfrak{so}(1,4)/\mathfrak{so}(2,3) \subset
\mathfrak{im}(4)$ algebra
\be\nno %
  [ \mathbf{P}^\pm_\mu, \mathbf{P}^\pm_\nu ] =\pm l^{-2} \mathbf{L}_{\mu\nu},~~
  {[} \mathbf{L}_{\mu\nu},\mathbf{P}^\pm_\ka {]} =
    \eta_{\nu\ka} \mathbf{P}^\pm_\mu - \eta_{\mu\ka} \mathbf{P}^\pm_\nu,\\\label{eq:so14/23}
  {[} \mathbf{L}_{\mu\nu},\mathbf{L}_{\ka\la} {]} =
    \eta_{\nu\ka} \mathbf{L}_{\mu\la}
  - \eta_{\nu\la} \mathbf{L}_{\mu\ka}
  + \eta_{\mu\la} \mathbf{L}_{\nu\ka}
  - \eta_{\mu\ka} \mathbf{L}_{\nu\la}.\quad
\ee%
Here the generators $\mathbf L^\pm_{\mu\nu}=\mathbf L_{\mu\nu}$ of
Lorentz algebra $\mathfrak{so}(1,3)$
 are the
same as that of the Poincar\'e algebra (\ref{eq:iso13}).

Thus, this confirms that on the \dS/\AdS\ spacetime of
possitive/negative constant curvature of radius $l$, there are
indeed other two  kinds of the Beltrami inertial frames with
inertial motions for 
the \dS/\AdS\ special relativity, 
respectively\cite{Lu,LZG, Hua,psr, BdS, TdS,IWR,NH,Lu05, C3,Yan,
OoI, PoI, Lu80}.


\section{
Three Kinds of Special Relativity as a Triple}\label{sec: SRT}

It is important to notice  from our consideration in last section
that for the Lorentz algebra $\mathfrak{so}(1,3)$  of \dS/\AdS\
algebra $\mathfrak{so}(1,4)/\mathfrak{so}(2,3)$ (\ref{eq:so14/23}),
the generators (\ref{eq:Lorentz'}) are the same as that of the
Poincar\'e algebra $\mathfrak{iso}(1,3)$ (\ref{eq:iso13}). This
would mean that there should be certain relations among
 $\mathfrak{iso}(1,3)/\mathfrak{so}(1,4)/\mathfrak{so}(2,3)\subset \mathfrak{im}_L(4)$
for three kinds of \SR, if their symmetries are subsets of
$\mathfrak{im}_L(4)$ at the same time. Let us  consider this
 issue further.

\subsection{The Special Relativity Triple with  Isotopy  Lorentz Group}

Firstly, it is also important to notice that in the space of
$\mathfrak{im}_L(4)$, not only the isotropy algebra
$\mathfrak{so}(1,3)$ is common, but the translation's generators
$\mathbf P _\mu$ of $\mathfrak{iso}(1,3)$ in (\ref{eq:generator})
can be directly given by the generators $\mathbf P^\pm _\mu$ of
$\mathfrak{so}(1,4)/\mathfrak{so}(2,3)$ in (\ref{eq:generatorpm}):
\be\label{eq: mathbf P}%
    {\mathbf P}_\mu :=\d 1 2 \left({\mathbf P}^+_\mu+{\mathbf P}^-_\mu \right) =
    \partial_\mu.
 \ee%
Then, the Poincar\'e algebra $\mathfrak{iso}(1,3)$
(\ref{eq:iso13}) can  be given by    \dS-algebra
$\mathfrak{so}(1,4)$ and \AdS-algebra $\mathfrak{so}(2,3)$ in
(\ref{eq:so14/23}) with
 same $l$. Thus, for three kinds of inertial
 transformations of
 Poincar\'e/\dS/\AdS-invariant \SR, the total number of generators  is only fourteen.
 In
 ${\cal T}_L$,
 there are still  other ten generators.
 In order to get the entire algebra   $\mathfrak{im}_L(4)$,
we may take the infinitesimal transformations and their generators
(\ref{eq: gntrs}) or we may also equivalently start with
  $\mathbf P^\pm_\mu$ in
 \dS/\AdS-algebra as the subalgebras of  $\mathfrak{im}_L(4)$ to get it:
 \be%
 [\mathbf P^- _\mu, \mathbf P^+_\nu ] =\begin{cases}
-l^{-2}\mathbf R_{\mu\nu} & \text{$\mu \neq \nu$}, \\
2 l^{-2}
\eta_{\mu(\nu)} \mathbf M_{(\nu)} & \text{ $\mu=\nu$},
\end{cases}%
\ee%
where no summation for repeated indexes in brackets and %
 \begin{equation}\label{eq:gnrtsrm}%
\begin{split}
\mathbf R_{\mu\nu} &= x_\mu \partial_\nu + x_\nu \partial_\mu \, ,
\qquad \mu \neq \nu \, , \\
\mathbf M_{\mu} =& - \left( x^{(\mu)} \r_{(\mu)} + \sum_\ka x^\ka
\r_\ka \right).
\end{split}\end{equation}
 It is straightforward to check
 that  twenty four generators
 $({\mathbf P}^\pm_\mu, {\mathbf L}_{\mu\nu}, {\mathbf R}_{\mu\nu}, {\mathbf M}_{\mu})$
    form the inertial motion algebra
   $\mathfrak{im}_L(4)$. In addition to the \dS/\AdS-algebraic relations
   (\ref{eq:so14/23}), the non-vanishing relations include
\begin{equation}
\begin{split}
 [ \mathbf P^\pm_{\mu}, \mathbf M_{\nu} ] = - \mathbf P^\mp_{\mu} -
 \delta_{\mu(\nu)}\mathbf P^\mp_{(\nu)},\quad
 [\mathbf{P}^\mp_\mu , \mathbf{R}_{\rho\sigma}]
  = \eta_{\mu\rho} \, \mathbf{P}^\pm_\sigma + \eta_{\mu\sigma} \, \mathbf{P}^\pm_\rho
  \qquad\\
\qquad[\mathbf{L}_{\mu\nu}, \mathbf{M}_{\rho}]
=\delta_{\mu(\rho)}\mathbf{R}_{\nu(\rho)}-
\delta_{\nu(\rho)}\mathbf{R}_{\mu(\rho)}, \hspace{3cm}\\
\hspace{1cm}[\mathbf{L}_{\mu\nu}, \mathbf{R}_{\rho\sigma} ] =
\begin{cases}
2\eta_{\mu\mu}\eta_{\nu\nu}(\mathbf{M}_{(\mu)} -\mathbf{M}_{(\nu)})
\quad
{\rm if}~ \mu=\rho, \nu=\sigma, (\mu=\sigma, \nu=\rho), \\
\eta_{\nu\rho}\mathbf{R}_{\mu\sigma} +
\eta_{\nu\sigma}\mathbf{R}_{\mu\rho}
-\eta_{\mu\rho}\mathbf{R}_{\nu\sigma}-\eta_{\mu\sigma}\mathbf{R}_{\nu\rho},~~{\rm
itc}.
\end{cases}
\end{split}\label{eq: im4}%
\end{equation}
It is clear that in the full algebra $\mathfrak{im}_L(4)$   there
are the  \dS, \AdS\ and Poincar\'e algebras as subalgebras
generated
  by  $({\mathbf P}^\pm_\mu, {\mathbf L}_{\mu\nu})$ and their combination
  $({\mathbf P}_\mu, {\mathbf L}_{\mu\nu})$
    for
  three kinds of \SR, respectively,  and the rest generators  $({\mathbf R}_{\mu\nu}, {\mathbf M}_{\mu})$
  do combine  three kinds of \SR\ as a whole. In fact,
  four generators $ \mathbf M_{\mu}$ form a Cartan
 subalgebra  $\mathfrak{h}$ that
 exchange the `translations' $ \mathbf P^\pm_{\mu}$ in \dS/\AdS-algebra from each other and  
also transform $\mathbf P_\mu$ of $ \mathfrak{iso}(1,3)$ among
themselves%
\be%
[ \mathbf P_{\mu}, \mathbf M_{\nu} ] = - \mathbf P_{\mu} -
\delta_{\mu(\nu)}\mathbf P_{(\nu)}.%
\ee%
 And the rest six generators $\mathbf R_{\mu\nu}$ play similar
roles: transforming  $\mathbf P_\mu^\pm$ of \dS/\AdS-algebra from
each other and  $\mathbf P_\mu$ of $\mathfrak{iso}(1,3)$ among
themselves%
\be%
  [\mathbf{P}_\mu , \mathbf{R}_{\rho\sigma}]
  = \eta_{\mu\rho} \, \mathbf{P}_\sigma + \eta_{\mu\sigma} \, \mathbf{P}_\rho
  .%
  \ee%

Thus, in addition to having a common Lorentz group, three kinds of
\SR\ act as a whole within $IM_L(4)$. It is called {\it the \SR\
triple}.

From  Eq. (\ref{eq: mathbf P}), it is natural to get
the other linear combination%
\be\label{eq: mathbf P'}%
    {\bf P}'_\mu :=\d 1 2 \left({\bf P}^+_\mu-{\mathbf P}^-_\mu
    \right)=-
l^{-2} x_\mu \sum x^\ka \r_\ka.
 \ee%
Actually, the generators $\bf P'_\mu, {\mathbf L}_{\mu\nu}$  form
the other  $\mathfrak{iso}(1,3)$ algebra isomorphic to the
Poincar\'e algebra (\ref{eq:iso13}). This algebra may be called the
second Poincar\'e algebra. It can be proved that this algebra
preserves the light-cone at the origin.

It is clear that in the Lie algebraic space of $\mathfrak{im}_L(4)$,
the set of generators of the \dS/\AdS\ algebra and two sets of
generators of the Poincar\'e algebras can be transferred from each
other by linear combinations between the generators as bases.

\subsection{On Other Symmetries in the Special Relativity Triple}

There are very rich substructures in the \SR\ triple. Let us
consider some of them at Lie algebraic level.

It is easy to check that sixteen generators $({\mathbf
L}_{\mu\nu}, \mathbf R_{\mu\nu}, \mathbf M_{\mu})$ form a
subalgebra $\mathfrak{gl}(4)$. Although there is Lorentz algebra
for isotropy, there are no generators for the transformation of
the origin to other points. Actually, if in such kind subalgebras
of $\mathfrak{im}_L(4)$ that there are no generators for
transitivity, which may be related to parameters $b^\mu$ in
(\ref{eq:LFT}), this kind of subalgebras may not be related to the
spacetime physics.
 Among  this kind of subalgebras, the second Poincar\'e algebra just mentioned
  is an interesting one, which keeps the light cone at the origin.
Therefore, for the spacetime physics, it is concerned with not
only the subsets of the ${\cal T}_L$, but also those subsets with
transitivity for all points in the frames. It is clear that the
\SR\ triple should contain all these subsets.

If we further consider within the \SR\ triple the algebras of ten
generators with a common space rotation algebra
$\mathfrak{so}(3)\subset \mathfrak{so}(1,3)$ as the isotropy
algebra rather than the Lorentz algebra $\mathfrak{so}(1,3)$, it
may follow all algebras of possible kinematics other than
Poincar\'e/\dS/\AdS-algebra, such as Newton-Hooke algebras
$\mathfrak{n}^\pm$, Galilei algebra, para-Poincar\'e algebra,
para-Galilei algebra as well as
 Carroll and
static algebras \cite{Bacry}. In addition,  their space-times may
follow, too. This may be seen from algebraic analysis including the
root system of $\mathfrak{im}(4)$. It is worthwhile to mention that
all these kinematics  should be based on the \PoR\ and appear as
subalgebras of $\mathfrak{im}(4)$ rather than by contractions under
different limits. This is different from the
`deformation/contraction' approach \cite{Bacry,IW}.

In addition, there are also very rich discrete symmetries for the
inertial motion transformation group ${IM}_L(4)$. Related to the 4-d
spacetime physics, the $CPT$ and so on should be considered.


\section{Concluding Remarks}

 We have  shown that based on the \PoR\ and the postulate for universal invariant
constants, three kinds of \SR\ form  the \SR\ triple with common
isotropy Lorentz group under  the \UWFL transformations ${\cal
T}_L$.

Since the inertial motion algebra $\mathfrak{im}(4)$ is very closely
related to the algebra  $\mathfrak{rp}(4)$
 of the 4-d real projective group $RP(4)$ on
4-d real projective space $\mathbb{RP}^4$, the projective geometry
approach may be applied (see, e.g., \cite{psr, Hua}) not only
locally but also globally. However, care should be taken, since
$\mathbb{RP}^4$ is not orientable. In our previous approach to the
\dS/\AdS\ \SR, in order to preserve the orientations the antipodal
identifications should not be taken (see, e.g.\cite{BdS, Lu80}).
This is also the case for the \SR\ triple. Actually, the
\Mink/\dS/\AdS-triple may also be implied in view of  algebraic
geometry. In fact, two quadratic forms ${\cal H}_\pm$  the
 \dS/\AdS-hyperboloid (\ref{eq:hyperboloids})
  share a common
 quadratic form $\eta_{\mu\nu}\xi^\mu \xi^\nu$, which can be given by the sum
 of quadratic forms  of
 ${\cal H}_\pm$,  invariant under
 Poincar\'e transformations.
In this Letter, we have mainly concerned with the local properties.
The global aspects are of course important issues.

All these approaches may be applied to three kinds of classical
geometries. For instance, if the isotropy algebra
$\mathfrak{so}(1,3)$ is replaced by an $\mathfrak{so}(4)$ algebra,
the algebra
$\mathfrak{iso}(4)/\mathfrak{so}(5)/\mathfrak{so}(1,4)$ for  4-d
Euclid/Riemann/Lobachevsky geometry  may follows, respectively.
The rest generators of  $\mathfrak{rp}(4)$ also transform among
themselves so that there is  a Euclid-Riemann-Lobachevsky-triple
of three kinds of classical geometries.
 In this case, there are also rich sub-geometry structures.

It should be emphasized that the  algebra of twenty four generators
can also be regarded as an entire `conformal' algebra for three
kinds of \SR\ or three kinds of classical geometry of one dimension
lower. This can be seen partially by the boundary of a 4-d
\AdS-spacetome as the conformal extensions of 3-d
\Mink/\dS/\AdS-spacetime, respectively \cite{C3}.

All these issues can be extended to any dimensions.

We have only considered the kinematic aspects of the \SR\ triple,
the dynamic aspects are of course very important, for which there
are lots of issues to be studied.

Although all these kinematics are based on the \PoR\ with inertial
frames involved, the algebraic combinatory approach may also be
employed for other kinds of coordinate frames. For instance, in
the \SR\ triple, the proper time coordinate may be taken. Thus, it
follows the triple for the \Mink-space, the Robertson-Walker
($RW$)-like \dS-space with an accelerated expanding 3-sphere  and
the $RW$-like \AdS-space with an oscillating
3-pseudosphere\cite{PoI,Lu80}. Our universe should prefer the
$RW$-like \dS-cosmos as its fate .

For the \SR\ triple, there is no gravity at all. In order to
describe gravity, the principle of localization of \SR\
\cite{duality, PoI, Lu80} should be applied to the \SR\ triple. In
our universe, the \SR\ triple should be reduced to the \dS\ \SR\
and its $RW$-like \dS-cosmos. Once gravity could be introduced,
this might be closely related to the reduction of the structure
group $IM(4)$ to its \dS\ sub-group as structure group. How to
realize the localization, symmetry breaking or reduction are also
important issues.



\begin{acknowledgments}
The authors would like to thank Profs/Drs  Q.K. Lu, Z. Chang, C.-G.
Huang, R.Q. Liu, J.Z. Pan, X.A. Ren, Y. Tian, S.K. Wang, K. Wu, X.N.
Wu, Z. Xu, Z.J. Zheng, and C.J. Zhu for valuable discussions. This
work is partly supported by NSFC (under Grants Nos.  90503002,
10701081,  10505004, 10675019), NKBRPC(2004CB318000) and Beijing
Jiao-Wei Key project (KZ200810028013).
\end{acknowledgments}

\end{document}